\documentclass[twocolumn,showpacs,aps,amsmath,amssymb]{revtex4}

\usepackage{graphicx}
\usepackage{dcolumn}
\usepackage{bm}
\usepackage{color}

\begin{document}


\newcommand{\be}{\begin{equation}} 
\newcommand{\ee}{\end{equation}} 
\newcommand{\ba}{\begin{eqnarray}} 
\newcommand{\ea}{\end{eqnarray}} 
\newcommand{\etal}{{\em et al.}}

\title{Investigation of the $^6$He cluster structures}

\author{L. Giot}
 \altaffiliation[Present address: ]{GSI, D-64291 Darmstadt, Germany}
\email{l.giot@gsi.de} 
\author{P. Roussel-Chomaz}
\author{C.E. Demonchy}
\author{W. Mittig}
\author{H. Savajols}
\affiliation{%
GANIL, BP 5027, 14076 Caen, France}
\author{N. Alamanos}
\author{F. Auger}
\author{A. Gillibert}
\author{C. Jouanne}
\author{V. Lapoux}
\author{L. Nalpas}
\author{E.C. Pollacco}
\author{J. L. Sida}
\author{F. Skaza}
\affiliation{
DSM/Dapnia CEA Saclay, 91191 Gif-sur-Yvette,France}
\author{M.D. Cortina-Gil}
\author{J. Fernandez-Vasquez}
\affiliation{
Universidad de Santiago de Compostela, 15706 Santiago de Compostela,
Spain}
\author{R.S. Mackintosh}
\affiliation{
Department of Physics and Astronomy, The Open University, Milton Keynes, 
MK7 6AA United Kingdom}
\author{A. Pakou}
\affiliation{
University of Ioannina, 45110 Ioannina, Greece}
\author{S. Pita}
\affiliation{
PCC, College de France, 11 place Marcelin Berthelot, 
75231 Paris, France}
\author{A. Rodin}
\author{S. Stepantsov}
\author{G. M. Ter Akopian}
\affiliation{
Flerov Laboratory of Nuclear Reactions, JINR, Dubna 141980, Russia}
\author{K. Rusek}
\affiliation{Department of Nuclear Reactions,
The Andrzej~Soltan~Institute~for~Nuclear Studies, \\
Hoza 69, PL 00-681 Warsaw, Poland}
\author{I. J. Thompson}
\affiliation{University of Surrey, Guildford,GU2 7XH, United Kingdom}
\author{R. Wolski}
\affiliation{The Henryk Niewodnicza\'nski Institute of Nuclear Physics PAN, \\
Radzikowskiego 152,
 PL-31-342 Cracow, Poland}

\date{\today}%

\begin{abstract}
The $\alpha$+2$\it{n}$ and $\it{t}$+$\it{t}$ clustering of the $^6$He ground 
state were 
investigated by means of the transfer reaction $^6$He($\it{p}$,$\it{t}$)$^4$He 
at 25 MeV/nucleon. 
The experiment was performed in inverse kinematics at GANIL with the SPEG spectrometer 
coupled to the MUST array.
Experimental data for the transfer reaction were analyzed by a DWBA calculation
including the two neutrons and the triton transfer. The couplings to the
$^6$He $\rightarrow$ $^4$He + 2$\it{n}$ breakup channels 
were taken into account with a
polarization potential deduced from a coupled-discretized-continuum channels analysis 
of the $^6$He+$^1$H 
elastic scattering measured at the same time. 
The influence on the calculations of the $\alpha$+$\it{t}$ 
exit potential and of the triton 
sequential transfer is discussed. 
The final calculation gives a spectroscopic factor close to one 
for the $\alpha$+2$\it{n}$ 
configuration as expected. The spectroscopic factor obtained for 
the $\it{t}$+$\it{t}$ configuration is much smaller than the theoretical predictions. 
\end{abstract}


\maketitle

\section{Introduction}
%
In the vicinity of the neutron dripline, the weak binding of the nuclei
leads to exotic features such as halos \cite{Hans95}. Furthermore, the two neutrons
halo systems, such as $^{6}$He, $^{11}$Li and $^{14}$Be, 
exhibit Borromean characteristics, whereby the two-body subsystems 
are unbound \cite{Zhuk93}.
Among these Borromean nuclei, the case of the $^{6}$He nucleus is of a special interest
from both experimental and theoretical points of view to study the 
halo phenomenon and 
3-body correlations,
especially because the $\alpha$ core can be represented as structureless. 
The importance of the 
dineutron and cigarlike $\alpha$+2$\it{n}$ configurations predicted for the 
$^{6}$He ground-state wave function \cite{Zhuk93} was investigated by means of
a 2$\it{n}$ transfer reaction or a radiative proton capture \cite{Ter98,Sauv01}. 
These two experiments give opposite conclusions concerning the relative importance 
of dineutron and cigarlike configurations.

An additional question arises whether the only contributions 
to the $^{6}$He ground-state wave function
are the cigar and di-neutron configurations, or if some $\it{t}$+$\it{t}$ 
clustering
is also present.
According to translational invariant shell model calculations \cite{Smir77}, 
the $^{6}$He nucleus
is expected to have a large spectroscopic amplitude for the $\it{t}$+$\it{t}$  
configuration as well as for the $\alpha$+2$\it{n}$ configuration, 
like the $^{6}$Li nucleus for the configurations $\alpha$+$\it{d}$ and 
$^{3}$He+$\it{t}$.
Microscopic multicluster calculations 
show also that the binding energy of $^{6}$He is better reproduced by including 
some $\it{t}$+$\it{t}$ clustering in the ground state wave function \cite{Csot93}. 
The spectroscopic factors
predicted for the $\it{t}$+$\it{t}$ configuration by the different models 
range from 0.44 \cite{Arai99} to 1.77 \cite{Smir77}.

Experimentally, this $\it{t}$+$\it{t}$ clustering was investigated for the first time 
by means of ($\it{t}$,$^{6}$He) transfer reactions on several targets and 
a spectroscopic factor S$_{t-t}$ of 1.77 was proposed \cite{Clar92}.
Recently, Wang {\em et al.} determined the charge radius of the $^6$He nucleus by using 
a laser spectroscopy method and compared the experimental extracted radius 
with those predicted by nuclear structure calculations \cite{Wang04}. 
The experimental radius is very close to the predictions of the 
$\alpha$+2$\it{n}$ cluster models but far away from those using a combination 
of $\alpha$+2$\it{n}$ 
and $\it{t}$+$\it{t}$ clusters.
To determine the importance of the $\alpha$+2$\it{n}$ and $\it{t}$+$\it{t}$ 
configurations of the $^6$He nucleus,
 Wolski {\em et al.} measured at Dubna the intermediate angles 
of the $^{6}$He($\it{p}$,$\it{t}$)$^{4}$He angular distribution \cite{Wols99}, in analogy with the 
reaction $^6$Li($\it{p}$,$^3$He)$^4$He used to study the relative importance of
the configurations $\alpha$+$\it{d}$ and $^{3}$He+$\it{t}$ for the $^{6}$Li nucleus 
\cite{Werb73}.
This reaction performed at 150 MeV can proceed as the transfer of two neutrons
 or of a $^{3}$H from the $^{6}$He nucleus.
Several analyses leading to very different spectroscopic factors S$_{t-t}$
 were performed on these experimental data within the distorted wave Born
 approximation (DWBA) framework. The DWBA analysis done by Wolski {\em et al.}
 suggests a spectroscopic factor S$_{t-t}$ of 0.42. Rusek {\em et al.} also 
 conclude to a small spectroscopic factor S$_{t-t}$ equal to 0.25 \cite{Ruse01}.
 The extreme case is the analysis of Oganessian {\em et al.},
  which described the data in a 
four-body three dimensional DWBA approach without any $\it{t}$+$\it{t}$ 
clustering \cite{Ogan99}.
At the opposite, Heiberg-Andersen {\em et al.} reproduced 
these data with a spectroscopic factor S$_{t-t}$ of 1.21 and took also 
into account the 2$\it{n}$ 
sequential transfer \cite{Heib01}. 
All these calculations call for data covering a wider range, 
especially at backward angles where the dependence of the $^{6}$He($\it{p}$,$\it{t}$)$^{4}$He differential
cross section with the value of the spectroscopic factor S$_{t-t}$ is important 
whereas the effect of the spectroscopic factor S$_{\alpha-2n}$ is dominant at forward angles. 
Hence to clarify the situation between the different 
values of the spectroscopic factor S$_{t-t}$, we measured at GANIL 
the complete angular distribution for the $^{6}$He($\it{p}$,$\it{t}$)$^{4}$He 
 with a special emphasis on the forward and backward angles
 which were never measured.
%
\section{Experimental method}
\begin{figure}[h]
\begin{center}
\includegraphics[width=8cm]{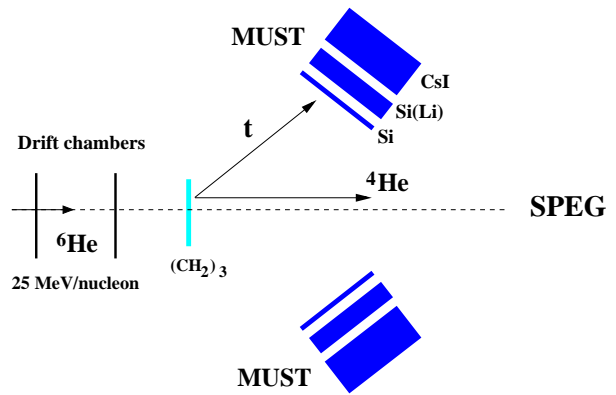}%
\caption{\label{fig:sketch} (Color online) Experimental setup.}
\end{center} 
\end{figure}

%
The experiment was carried out at the GANIL coupled cyclotron facility. 
The composite secondary beam was produced by the fragmentation of a 780 MeV
$^{13}$C beam of 5 $\mu$Ae on a 1040 mg/cm$^2$ carbon production target 
located between the two superconducting solenoids of the SISSI device \cite{Ann97}.
The $^{6}$He nuclei were selected with the two dipoles of the $\alpha$ spectrometer 
and an achromatic Al degrader located at the dispersive plane between these two dipoles.
The only contaminant was $^{9}$Be at a level of 1\%.
The resulting 150 MeV $^{6}$He beam with an average intensity of 1.1 10$^{5}$ pps 
impinged on a (CH$_2$)$_3$ target, 18 mg/cm$^2$ thick, located in the reaction chamber.
 A sketch of the experimental setup
 is shown on Fig. \ref{fig:sketch}.
Due to the large emittance of the secondary fragmentation beam,
the incident angle and the position on the target of the nuclei were monitored 
event by event by two low pressure drift chambers \cite{Mac98}.
The angular and position resolutions on the target were respectively 
0.14$^\circ$ and 2.4 mm.

\begin{figure}[h] 
\begin{center}
\includegraphics[width=7cm,angle=0]{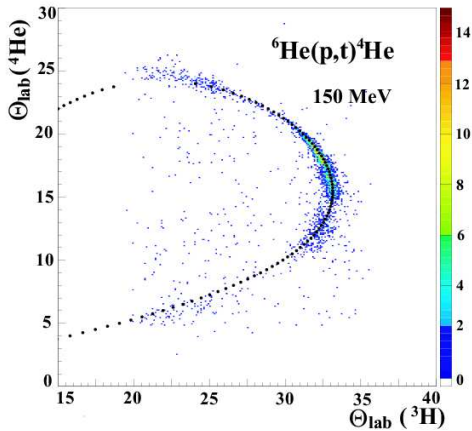}%
\caption{\label{fig:must} (Color online) Scatterplot of the $^{4}$He laboratory angle versus 
the $^{3}$H laboratory
angle detected in coincidence in the MUST array for the 
$^6$He($\it{p}$,$\it{t}$)$^4$He 
reaction.}
\end{center}
\end{figure}

The $^4$He and $^3$H from the $^6$He($\it{p}$,$\it{t}$)$^4$He reaction for center of mass angles 
between 20$^\circ$ and 110$^\circ$ were detected in coincidence by the eight telescopes of
MUST silicon detector array \cite{Blum99}. These detectors were separated 
in two groups
arranged to form two 12x12 cm$^{2}$ squares placed on each side of the beam, one covering an angular range 
between 6$^\circ$ and 24$^\circ$
 and the other one between 20$^\circ$ and 38$^\circ$ with respect to the beam direction.
 The angular coverage in the vertical direction was $\pm$ 9$^\circ$.
 Each of the MUST telescopes is composed of a doubled-sided silicon strip detector 
 backed by a Si(Li) and a CsI scintillator which all give an energy measurement. 
The silicon strip detector is 300 $\mu$m thick with 60 strips (1 mm wide) on each side 
and provides X-Y position measurement, from which the scattering angle is determined.
The energy resolution in the silicon strips detectors was 65 keV and the angular
resolution 0.15$^\circ$. 
The $^{4}$He and $^3$H nuclei were identified by the standard $\Delta$E-E technique.
Due to a saturation of the Si(Li) preamplifiers, we did not measure the total energy of 
the $\alpha$ particles. The off-line identification combined with a gate on the 
 kinematical locus of the transfer reaction on the $^3$H plot, energy 
 versus scattering angle, reduced the background from the carbon component of the target. 
 Fig. \ref{fig:must} shows the angular correlation of the $^4$He and $^3$H nuclei.
 The dotted line corresponds to the calculated 2-body kinematical line for the
 ($\it{p}$,$\it{t}$) transfer 
 reaction towards the $^6$He ground state. 
 The remaining backgound in the vicinity of this line
was evaluated and subtracted to the region of interest.

The forward and backward center of mass angles of the angular distribution for the 
$^6$He($\it{p}$,$\it{t}$)$^4$He reaction were measured with the SPEG spectrometer
 \cite{Bia89} by detecting respectively the high energy $^4$He
 and the high energy triton at forward laboratory angles. 
 The particles were identified in the focal plane by the energy loss
measured in an ionization chamber and the residual energy measured in plastic
scintillators. The momentum and the scattering angle were obtained by
track reconstruction of the trajectory as determined by two drift chambers located near
the focal plane of the spectrometer.
The spectrometer was also used to measure the elastic scattering 
$^{6}$He($\it{p}$,$\it{p}$)$^{6}$He angular distribution 
from 14$^\circ$ c.m. to 60$^\circ$ c.m. .

\begin{figure}[h]
\begin{center}
\includegraphics[width=7cm,angle=0]{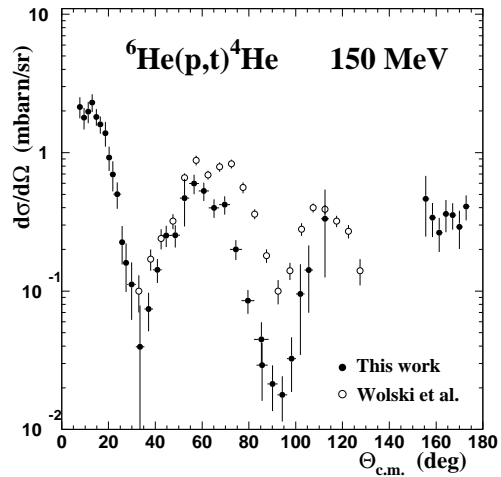}%
\caption{\label{fig:compadubna} Differential cross section 
for the 
$^6$He($\it{p}$,$\it{t}$)$^4$He reaction.
The full circles correspond to the present data. The open circles are the
data measured at Dubna \cite{Wols99}.}
\end{center}
\end{figure}

To extract differential cross sections, data were corrected for the geometrical efficiency
 of the detection in SPEG or MUST. 
This efficiency was determined through a Monte-Carlo simulation whose
ingredients are the detector geometry, their experimental angular and energy resolutions,
the position and the width of the beam on the target. 
The error on the MUST detection efficiency deduced from the Monte-Carlo simulation 
is estimated to be 5 \%.
The absolute normalisation for the elastic scattering data on the protons in the
(CH$_2$)$_3$ target and for the transfer reaction was obtained from the elastic scattering 
on $^{12}$C which was measured simultaneously. Elastic scattering calculations
 for the system $^6$He+$^{12}$C using
different optical potentials  \cite{Alkh96,Brog81} 
 show that the angular distribution at forward angles up to 7$^\circ$ c.m. is
dominated by Coulomb scattering and is rather insensitive to the potential used.
Therefore the absolute normalisation of the data was obtained from the measured cross
section on the first maximum of the $^6$He($^{12}$C,$^{12}$C)$^6$He angular distribution. 
The uncertainty on the normalisation is of the order of 10 \%.
The same normalisation factor was applied to the transfer data measured with the SPEG spectrometer.
In the overlap domain between 19$^\circ$ c.m. and 27$^\circ$ c.m. for the transfer data 
obtained with SPEG or MUST, the agreement was good. The final differential cross section 
in this
angular region is the statistical average value between the two sets of data. 
Fig. \ref{fig:compadubna} displays the transfer data obtained in the present
experiment together with results measured previously 
at Dubna at the same energy \cite{Wols99}. The uncertainty of the dubna differential cross sections is
estimated to be within a 30\% limit, mainly due to the beam monitoring errors. 
The error bars displayed for the GANIL data are purely statistical.
The data points between 120$^\circ$ and 155$^\circ$ could not be obtained due to a
lack of statistics in a set of runs with the MUST array positionned in this angular region.
The main difference between the two sets of data obtained at GANIL and Dubna is related 
to the width of the second 
minimum around 90$^\circ$ which is larger in our case. 
The Dubna data were extracted from the energy correlation 
between the $\alpha$ particles and the tritons detected in coincidence in two telescopes.
This energy correlation presented a strong background, caused by the breakup of the $^{6}$He particles 
on the carbon component of the target. 
The breakup background is not uniform and could explain the difference observed between 
the two sets of data in the angular range corresponding to the deep minimum around 
90$^\circ$. 
Here, the angular correlation of the $^{4}$He and $^{3}$H nuclei combined 
with a selection on the kinematical locus on the $^3$H plot, energy versus scattering angle, 
reduced strongly the breakup background and
improved the quality of the data, as shown on Fig. \ref{fig:must}.
 Moreover, the oscillation widths of the GANIL data are in better agreement with the 
new data measured at Dubna by Stepantsov {\em et al.} \cite{Step03}.

\section{Elastic scattering and 
the $^{6}$H\lowercase{e} + $\it{\lowercase{p}}$ optical potential}
\begin{figure}[h]
\begin{center}
\includegraphics[width=7cm,angle=0]{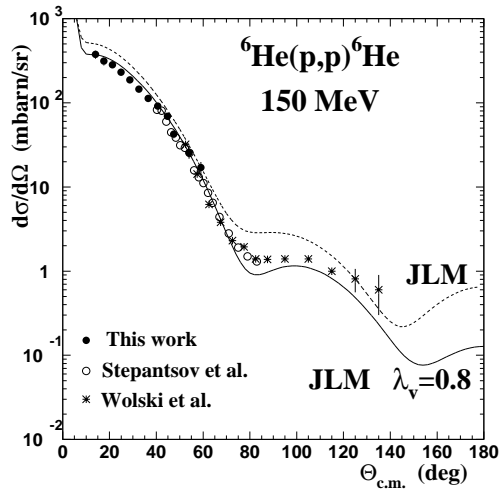}%
\caption{\label{fig:elasjlm} Differential cross section 
for the $^6$He($\it{p}$,$\it{p}$)$^6$He 
elastic scattering compared to JLM calculations.
The full circles correspond to the present data. The open circles and the crosses
 are the
data measured at Dubna \cite{Wols99, Step02}.}
\end{center}
\end{figure}

%
%
The $^{6}$He + $\it{p}$ elastic scattering data obtained here complement the data 
measured at Dubna at the same energy in two differents runs \cite{Wols99,Step02}.
Fig. \ref{fig:elasjlm} shows the three sets of data which are in excellent agreement.
These new data allow to better determine the nuclear interaction potential 
$^{6}$He + $\it{p}$,
 which is an essential ingredient for the analysis of
the transfer reaction, since it is necessary for the entrance channel in the DWBA calculation.
Previous results obtained on $^{6}$He + $\it{p}$ elastic scattering in the same energy range
have shown that the nucleon-nucleus optical models potentials used for stable nuclei
have to be modified in the case of loosely bound nuclei such 
as $^{6}$He \cite{Lapo01,Ruse01}.
Coupling to the continuum and to the resonant states are expected 
to play a significant role since the scattering
states are much closer to the continuum states than in stable nuclei.
In this work, we tried two approaches to take into account these couplings to the continuum 
and thus the breakup effects. 
First, these effects were phenomenologically simulated by reducing the real part
of the potential or by adding a dynamical polarization 
potential as discussed in Ref. \cite{Lapo01}. In this case, 
the interaction potential $^{6}$He + $\it{p}$
was calculated within the approach derived by Jeukenne, Lejeune  and Mahaux 
(JLM) \cite{Jeuk77}. 
As seen in Fig. \ref{fig:elasjlm}, this renormalization of the real part of the JLM potential 
allows also to reproduce the sets of data obtained at GANIL and Dubna reasonably
well. 

\begin{figure}[htp]
\begin{center}
\includegraphics[width=6cm,angle=0]{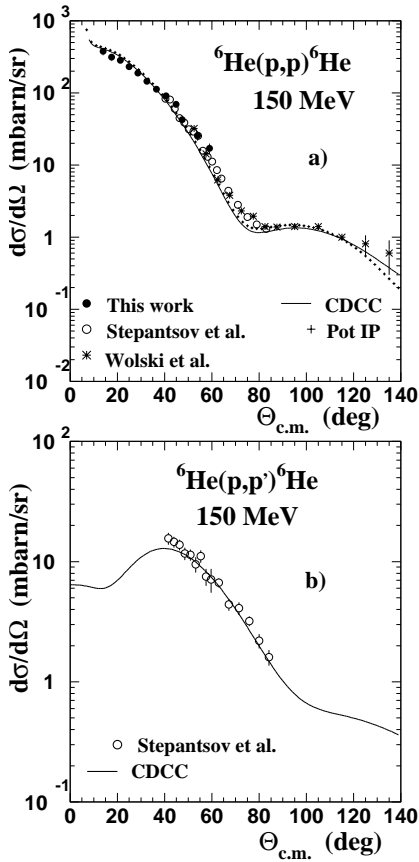}%
\caption{\label{fig:CDCC} a) CDCC calculations for the elastic scattering of $^6$He from
$^1$H. The experimental data are from this work (black circles) 
and from Ref. \cite{Wols99,Step02}. b) CDCC calculations for the inelastic scattering
of $^6$He from $^1$H leading to the 2$^+$ resonant state of $^6$He at excitation
energy of 1.8 MeV. The experimental data are from Ref. \cite{Step02}.}
\end{center}
\end{figure}

In a second approach, the coupling to the continuum was explicitly included by means of the
coupled-discretized-continuum-channels (CDCC) method.
This approach previously used by Rusek {\em et al.} \cite{Ruse01,Ruse00} assumes 
a two body
cluster model $\alpha$+2$\it{n}$ for the $^{6}$He nucleus with the 
spin of the 2$\it{n}$ cluster
 set to $\it{s}$=0. 
 The wave functions $\Psi$$_0$($\it{r}$) and $\Psi$$_2$($\it{r}$) describing 
 the relative motion 
of the two clusters in the $^{6}$He ground state and the $^{6}$He(2$^+$) resonant
 state 
were calculated in potential wells 
whose depths were varied to reproduce respectively the binding energy of 0.975 MeV 
and the excitation energy of 1.8 MeV. 
The parameters of these Woods-Saxon binding potentials are listed in Table
\ref{tab1:pot_bind}.
\begin{table*}[h]
\begin{ruledtabular}
\begin{tabular}{c c c c c c c c}
& & V$_0$ & R$_{0}$ & a$_0$ & Ref. & S$_a$ & Ref.\\
& & (MeV) & (fm) & (fm) &  \\
\hline
$^6$He$_{g.s.}$ = $\alpha$ + 2$\it{n}$ & & 69.393 & 1.9 & 0.65 & \cite{Ruse00}
 & 1. & This work \\
$^6$He$_{2+}$ = $\alpha$ + 2$\it{n}$ & & 80.427 & 1.9 & 0.65 & \cite{Ruse00}
 & 1. & \cite{Ruse00} \\
$^6$He$_{g.s.}$ = $\it{t}$ + $\it{t}$ & & 80. & 1.9 & 0.65 & \cite{Clar92} 
& -0.28 & This work \\
$^3$H$_{g.s.}$ = $\it{p}$ + 2$\it{n}$ & & 44.382  & 2. & 0.60 & \cite{Neud72} 
& 1. & \cite{Neme88} \\
$^4$He$_{g.s.}$ = $\it{p}$ + $\it{t}$ & & 61.4 & 2. & 0.60 & \cite{Neud72} 
& 1.4142 & \cite{Neme88} \\
$^4$He$_{g.s.}$ = $^3$He + $\it{n}$ & & 61.4 & 1.5 & 0.40 & This work 
& -1.4142 & \cite{Neme88} \\
\end{tabular}
\end{ruledtabular}
\caption{Parameters of the binding potentials and the spectroscopic amplitudes.}
\label{tab1:pot_bind}
\end{table*}
 The continuum above the $^{6}$He $\rightarrow$ $\alpha$+2$\it{n}$ breakup
 threshold was discretized into a series of momentum bins with respect 
 to the relative $\alpha$-2$\it{n}$ momentum $\it{k}$. The lowest bins were of 
 $\Delta$$\it{k}$=0.25 fm$^{-1}$ while all the others were of 
 $\Delta$$\it{k}$=0.2 fm$^{-1}$.
 The model space was truncated at the energies close to the 
 $\it{t}$+$\it{t}$ breakup threshold. 
 The wave function $\Psi$($\it{r}$) representing a bin is the average function over 
 the bin width
 of the cluster wave functions $\phi$($\it{r}$,$\it{k}$) in the bin,
\be
\Psi(r)=\frac{1}{\sqrt{N\Delta{k}}}\int_{\Delta k} \phi(r,k)dk
\label{eq1}
\ee
  where $\it{r}$ is the $^{4}$He-2$\it{n}$ distance 
  and $\it{N}$ is the normalization factor. 
 All the spectroscopic amplitudes for the couplings are assumed to be equal to one.
The central and coupling potentials $\it{V}$$_i$$_\rightarrow$$_f$($\it{R}$)
used in the CDCC calculations were derived from $^4$He-$\it{p}$ and 
2$\it{n}$-$\it{p}$ potentials 
by means of the single-folding method,
\begin{eqnarray}
V_{i\rightarrow f}(R)= &\langle \Psi_{f}(r)\vert U_{2n-p} (\vert \vec R + 2/3  \vec r
\vert ) \nonumber\\
  + &U_{^4He-p} (\vert \vec R - 1/3  \vec r \vert ) 
\vert \Psi_{i}(r) \rangle
\end{eqnarray}
where $\it{R}$ is the distance between the $^6$He nucleus and the proton. 
The 2$\it{n}$-$\it{p}$ potential
was assumed to be the same as for $\it{d}$-$\it{p}$. 
The parameters of the optical potentials $\it{U}$$_{2n-p}$ and $\it{U}$$_{^4He-p}$ 
, listed in Table \ref{tab2:pot_opt}, were obtained 
by fitting the elastic scattering data of deuterons and alpha particles from protons
at the required energy \cite{Hinte68,Bach72}. 
The results of the CDCC calculations for the $^6$He+$\it{p}$ elastic scattering, 
performed using version FRXY-1c 
of the code FRESCO \cite{Thom88}, are plotted in Fig. \ref{fig:CDCC}
together with the three sets of data measured at GANIL and Dubna.

These CDCC calculations reproduce, too, the values 
and the slope of the differential cross section 
 for the inelastic scattering exciting $^6$He to its 2$^+$ resonant state 
 \cite{Step02}.
The calculated value of the reduced transition 
probability $\it{B}$($\it{E2}$;g.s.$\rightarrow$2$^+$)=7.08 $\it{e}$$^2$ fm$^4$ 
is larger than the value of 3.21
$\it{e}$$^2$ fm$^4$ published earlier by Aumann {\em et al.} \cite{Auma99}.
However, it should be noticed that the determination of $\it{B}$($\it{E2}$) 
is strongly model dependent when the reaction is not dominated by the Coulomb interaction.
In particular, the value of $\it{B}$($\it{E2}$) depends on the choice 
of the neutron density distribution. 

By inversion from the elastic channel $\it{S}$ matrix, generated 
by the CDCC calculations, a local potential $^6$He+$\it{p}$ including 
the breakup effects,
 is obtained.
The inversion is carried out using the iterative-perturbative (IP) 
method \cite{Coop00,Mack03}. 
The angular distribution calculated within the CDCC approach or with the
local potential, named $^{\prime}$Pot. IP$^{\prime}$, can hardly be 
distinguished in Fig. \ref{fig:CDCC}. 
The reaction cross sections calculated 
respectively with the potential 
IP and the renormalized JLM potential are respectively $\sigma$$_R$ = 532 mb and 394 mb.
The reaction cross section for $^6$He on proton has been measured at 36 MeV/nucleon
using the transmission method and a value of $\sigma$$_R$ = 409$\pm$22 mb was obtained 
\cite{Devi01}.
Considering that in the present energy domain, 
the reaction cross section $\sigma$$_R$  
increases when the energy of the projectile decreases \cite{Kox87},
 the experimental value at 25 MeV/nucleon should be closer to the value calculated with 
the IP potential.
This IP local potential and the renormalized
JLM potential will be tested  in the next section as entrance potentials 
for the DWBA calculation of the $^{6}$He($\it{p}$,$\it{t}$)$^{4}$He.
%
%
\section{DWBA analysis of 
$^{6}$H\lowercase{e}($\it{\lowercase{p}}$,$\it{\lowercase{t}}$)$^{4}$H\lowercase{e}}
\begin{figure}[h]
\begin{center}
\includegraphics[width=6cm,angle=0]{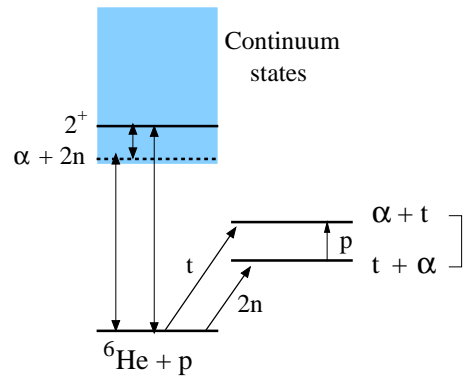}%
\caption{\label{fig:coupling} 
(Color online) Scheme of the $^6$He($\it{p}$,$\it{t}$)$^4$He DWBA calculation.}
\end{center}
\end{figure}

%
DWBA calculation including both 2$\it{n}$ and $\it{t}$ transfer 
from the $^6$He ground state
 were performed on the  
$^{6}$He($\it{p}$,$\it{t}$)$^{4}$He data with the code FRESCO used in its finite range option.
A sketch of the calculation is shown in Fig. \ref{fig:coupling}.
The couplings to the continuum states were taken into account with the entrance 
channel potential $^6$He+$\it{p}$ calculated, as described above within 
the JLM or the CDCC framework.
The effect of the triton sequential transfer ($\it{t}$=2$\it{n}$+$\it{p}$) 
was also investigated. 
The wave functions describing the relative motion of the two clusters 
$\alpha$+2$\it{n}$,  
$\it{t}$+$\it{t}$, $\it{p}$+$\it{t}$ and $\it{p}$+2$\it{n}$ 
respectively in the ground state of $^6$He, $^4$He and $^3$H
were obtained from Woods-Saxon potentials 
 with the well depth adjusted to reproduce the corresponding binding energies, according 
 to the usual separation energy prescription. The remnant potential 
 V$_{p-t}$ was taken from Ref. \cite{Neud72} 
and the remnant potential V$_{\alpha}$$_{-p}$ was determined in the previous section.
All the potentials and the spectroscopic amplitudes used in the calculation
are listed in Table \ref{tab1:pot_bind} and \ref{tab2:pot_opt}.

\begin{figure}[h]
\begin{center}
\includegraphics[width=8cm,angle=0]{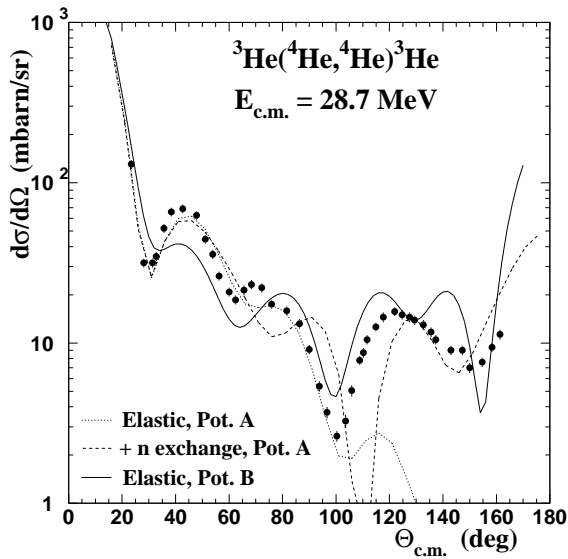}%
\caption{\label{fig:he3he4_pot} 
Differential cross section for the $^3$He($\alpha$,$\alpha$)$^3$He reaction
compared with DWBA calculations assuming only the elastic scattering and after 
the one neutron exchange. The experimental data are from Schwandt 
{\em et al.} \cite{Scha69}.}
\end{center}
\end{figure}

Special care was taken in the choice of the potential for 
the $\alpha$+$\it{t}$ exit channel
and several potentials were considered 
\cite{Wols99,Ogan99,Ruse01,Neud72,Timo01,Buck88}.
To obtain the $\alpha$+$\it{t}$ potential, we used $\alpha$+$^3$He elastic 
scattering data at E$_c$$_.$$_m$$_.$ = 28.7 MeV \cite{Scha69} 
as the $^3$H($\alpha$,$\alpha$)$^3$H reaction was not studied  
 in the energy range considered presently. Two approaches were considered 
 to extract the $\alpha$+$^3$He optical potential.
 The process of one neutron transfer, which is not distinguishable 
experimentally from the elastic scattering,  
 was first explicitly taken into account 
 in a DWBA analysis of the $^3$He($\alpha$,$\alpha$)$^3$He reaction.
 The dotted and dashed curves on Fig. \ref{fig:he3he4_pot} show respectively 
 the contribution of the elastic scattering and the one neutron exchange to
 the $^3$He($\alpha$,$\alpha$)$^3$He reaction. 
 Hence, the potential A extracted from the DWBA analysis represents only the 
 elastic scattering between the $\alpha$ and the $^3$He.
Next, we used the potential B obtained in Ref.\cite{Ruse01} fitted on 
the complete differential cross section of the
$^3$He($^4$He,$^4$He)$^3$He
elastic scattering.
This fit on the $\alpha$+$^3$He data corresponds to the
solid line on Fig. \ref{fig:he3he4_pot}.
Although the fit is still far from perfect,
it reproduces the gross features of the measured angular distribution.
 The values of these two exit potentials A and B are given 
 in Table \ref{tab2:pot_opt}. 

\begin{table*}[h]
\begin{ruledtabular}
\begin{tabular}{c c c c c c c c c c c}
& V$_0$ & R$_{0}$ & a$_0$ & W$_v$ & R$_v$ & a$_v$ 
& W$_s$ & R$_s$ & a$_s$ & Ref.\\
& (MeV) & (fm) & (fm) & (MeV) & (fm) & (fm) &
(MeV) & (fm) & (fm) &   \\
\hline
$\it{d}$ + $\it{p}$ & 65.80 & 1.25 & 0.501 & 0.00 & 0.00 & 0.00 & 10.00 
& 1.20 & 0.517 & \cite{Ruse01} \\
$^3$He + $^3$He & 60.00 & 2.58 & 0.417 & 0.00 & 0.00 & 0.00 & 0.00 
& 0.00 & 0.00 & This work \\
$\alpha$ + $\it{p}$ & 48.90 & 1.75 & 0.477 & 0.557 & 1.75 & 0.477 & 0.00 
& 0.00 & 0.00 & This work \\
$\alpha$ + $^3$He, Pot. A & 80.95 & 2.39 & 0.829 & 7.49 & 4.12 & 0.0196 & 0.00 
& 0.00 & 0.00 & This work \\
$\alpha$ + $^3$He, Pot. B & 142.92 & 2.57 & 0.271 & 0.86 & 6.88 & 0.972 & 0.00 
& 0.00 & 0.00 & \cite{Ruse01} \\
\end{tabular}
\end{ruledtabular}
\caption{Parameters of the input optical model potentials.}
\label{tab2:pot_opt}
\end{table*}

\begin{figure}[h]
\begin{center}
\includegraphics[width=8cm,angle=0]{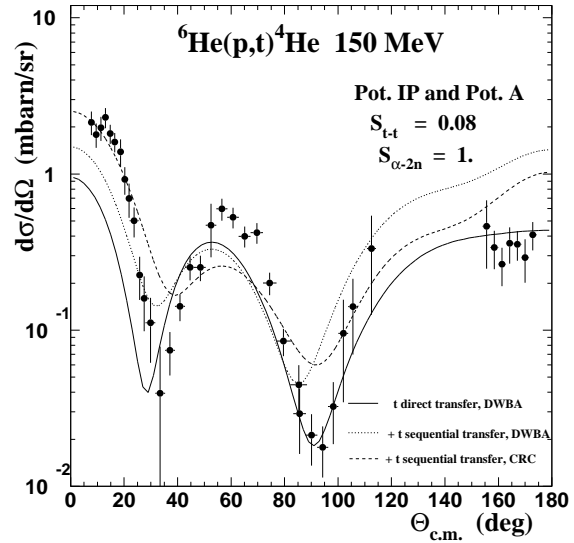}%
\caption{\label{fig:tsequent} Effect of the sequential transfer of the 
triton ($\it{t}$=2$\it{n}$+$\it{p}$)
on the $^{6}$He($\it{p}$,$\it{t}$)$^{4}$He reaction.}
\end{center}
\end{figure}

Both the direct and the sequential transfer of the triton were included in a 
DWBA calculation, described on Fig. \ref{fig:coupling}, and 
then in a coupled reaction channels calculation (CRC) where the backcouplings 
were taken into account.
The proton transfer, which is the second step of the triton sequential transfer, can be seen as a 
proton exchange between the two exit channels $^3$H+$\alpha$ and $\alpha$+$^3$H
and 
is directly included in the DWBA calculation. 
Hence, the exit channel potential $\alpha$+$^3$H has not to take into account the proton 
exchange but only the elastic scattering. 
This corresponds to the potential A, as determined above, for the exit channel.
A comparison between a DWBA calculation with only the triton direct transfer 
and DWBA and CRC calculations including the sequential transfer 
$\it{t}$=2$\it{n}$+$\it{p}$  
is shown on Fig. \ref{fig:tsequent}. These calculations do not allow to reproduce at the same
time the experimental data at forward and backward angles. 
Futhermore, the amplitude and/or the position of the oscillation 
at 60$^\circ$ is not reproduced.

\begin{figure}[h]
\begin{center}
\includegraphics[width=8cm,angle=0]{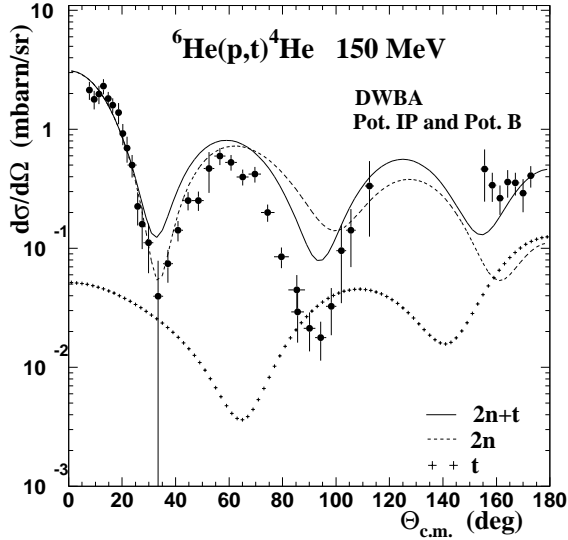}%
\caption{\label{fig:transdwba} Contributions of 2$\it{n}$ and $\it{t}$ transfer 
to the DWBA
 calculation of $^{6}$He($\it{p}$,$\it{t}$)$^{4}$He.}
\end{center}
\end{figure}

Hence, we used Potential B for the exit channel of the reaction, considering that 
the proton exchange in the $\alpha$+$\it{t}$ partition, 
and consequently the sequential triton transfer, 
could be in
some sense described by the exchange term included in the Potential B. 
Only the 2$\it{n}$ and the direct
triton transfer are thus explicitly included in the DWBA calculation. 
Fig. \ref{fig:transdwba} compares the $^{6}$He($\it{p}$,$\it{t}$)$^4$He differential cross section obtained at GANIL 
with the DWBA calculation using Potential IP for the entrance channel and Potential B for the exit channel, 
which gives actually the best description of the features of the $^{6}$He($\it{p}$,$\it{t}$)$^4$He angular distribution. 
The spectroscopic factors for the $\alpha$+2$\it{n}$ and $\it{t}$+$\it{t}$ 
configurations were
adjusted to reproduce the data. 
The adjustment on Fig. \ref{fig:transdwba} is a compromise between the experimental values 
at forward and backward angles and the width of the second oscillation.
The dashed line on Fig. \ref{fig:transdwba} 
corresponds to the DWBA calculation where only the 2$\it{n}$ transfer 
is taken into account 
with a
spectroscopic factor S$_{\alpha-2n}$ equal to 1, which is close to the predicted value.
The spectroscopic factor S$_{t-t}$, determined from the comparison with the data at backward angles, 
is equal to 0.08 with an uncertainty of the order of 50\%.
The crosses on Fig. \ref{fig:transdwba} correspond only to the triton transfer.
The solid line is the coherent sum of the two processes with these values of their
spectroscopic factors. 
Even if the spectroscopic factor of the $\it{t}$+$\it{t}$ clustering is small, 
the $\it{t}$-$\it{t}$ component 
is essential to reproduce the backward angles.

\begin{figure}[htp]
\begin{center}
\includegraphics[width=8cm,angle=0]{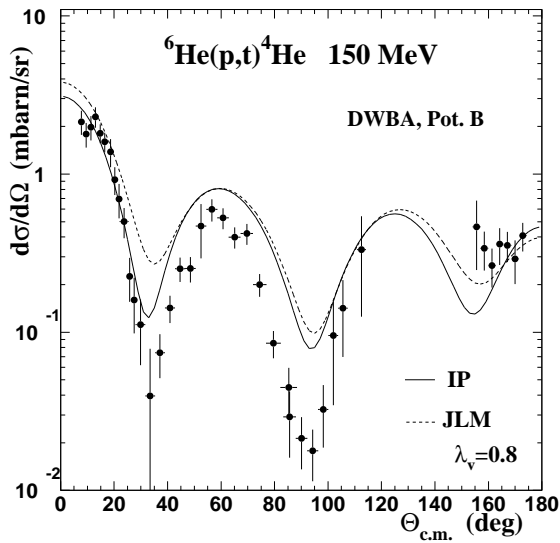}%
\caption{\label{fig:transdwbajlm} Dependence of the results of the DWBA calculation on the
choice of the $^6$He + $\it{p}$ entrance potential.}
\end{center}
\end{figure}

\begin{figure}[h]
\begin{center}
\includegraphics[width=8cm,angle=0]{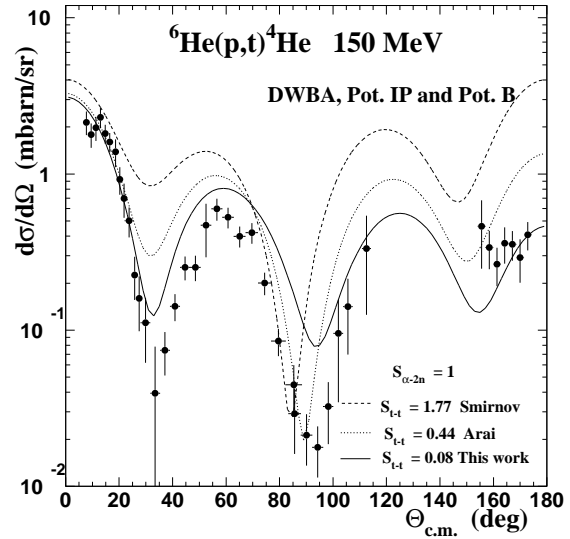}%
\caption{\label{fig:stt} Dependence of the DWBA calculation with the value of the
spectroscopic factor S$_{t-t}$
.}
\end{center}
\end{figure}

We tried to include the sequential transfer of the two neutrons, 
processing via the $^5$He+$\it{d}$ channel. 
This calculation required the optical potential for the 
$^5$He+$\it{d}$ system. Of course, no deuton elastic scattering data exist for this unbound nucleus.
A first potential was obtained from the $^6$He($\it{p}$,$\it{d}$)$^5$He 
reaction \cite{Wols99}. 
The second potential 
was calculated from a method proposed by Keaton {\em et al.} convoluting 
the potentials $^5$He+$\it{p}$ and
$^5$He+$\it{n}$, derived from the CH89 parametrization, 
with a deuteron wave function
 \cite{Keat73,Varn91,Laco80}. 
These two potentials provided a very poor reproduction for the one and two neutrons transfer reactions.
Therefore, we removed this channel in the final caculation, due to the lack of a reliable potential for the
unbound nucleus $^5$He.

Finally, both entrance potentials including breakup effects, obtained in the previous section,
were also tested in Fig. \ref{fig:transdwbajlm}. 
Obviously, the IP potential deduced from a coupled-discretized-continuum-channels calculation, 
where the couplings to the continuum are explicitly taken into account, 
improves the description of the data compared to the JLM approach. 

Fig. \ref{fig:stt} shows the dependence of the DWBA calculation with the value of the 
spectroscopic 
factor S$_{t-t}$ for 3 cases: $i)$ the value derived from the present work 
$ii)$ the value obtained in the 3-body cluster model by K. Arai {\em et al.} \cite{Arai99} 
and $iii)$ finally the value from a translational invariant shell model calculation \cite{Smir77}. 
It is clear that in the present analysis, the two last values strongly overestimate the cross section measured
at backward angles, and that the largest value even affects the reproduction of the most forward angles of
the data. The theoretical values are considerably outside the range of the uncertainty of the 
experimental data. The different approaches of analysis performed within this work showed that the final results 
somewhat depend on the potentials used in the calculation, especially in the exit channel potential. 
Considering the difficulties mentioned previously on this potential, one can not exclude that another type
of approach used to derive this potential could alter our conclusion.

\section{Conclusions}   
%
The $^{6}$He($\it{p}$,$\it{t}$)$^{4}$He reaction at 150 MeV has been investigated 
at GANIL in order 
to provide insight on the $^6$He cluster structure: the $\alpha$+2$\it{n}$ 
and di-triton configurations. 
The transfer differential cross section were 
measured with the SPEG spectrometer coupled to the MUST array. 
The transfer data obtained at forward and backward angles allowed 
to determine the spectroscopic factors 
S$_{\alpha-2n}$ and S$_{t-t}$ and thus the contribution of the 
configurations $\alpha$+2$\it{n}$ and $\it{t}$+$\it{t}$ 
to the $^6$He ground state wave function.
The $^{6}$He($\it{p}$,$\it{t}$)$^{4}$He data were analyzed
 by means 
of the DWBA and the coupled channels method taking into account the direct
2$\it{n}$, the direct triton transfer and also the sequential transfer 
 of the triton. 
 The $^6$He+$\it{p}$ entrance channel optical potential of these calculations
 was obtained by the inversion of the elastic channel $\it{S}$ 
 matrix generated
 from a CDCC calculation 
which took into account all the couplings to the continuum states.
This CDCC calculation, where the continuum above the $\alpha$+2$\it{n}$ threshold  
is discretized, reproduced the elastic 
$^6$He($\it{p}$,$\it{p}$)$^6$He, also measured in this experiment,
and the $^6$He($\it{p}$,$\it{p}$')$^6$He inelastic 
data available at the same energy as the transfer reaction.
A detailed study of the exit channel was performed. 
The difficulties encountered in this part of 
the analysis are related to the lack of $^4$He+$\it{t}$ 
elastic scattering data in the energy 
range considered presently, and to the strong effects of the neutron exchange 
in the $^4$He+$^3$He 
system which was used instead. 
The best description of the experimental data was obtained with a 
DWBA calculation taking into 
account the 2$\it{n}$ and $\it{t}$ direct transfer. 
The triton sequential transfer is assumed to be directly 
included in 
the exchange term of the exit potential $\alpha$+$^3$H. 
The present work shows that the DWBA analysis of the transfer data 
is strongly dependent 
on the chosen potentials. Nethertheless, the transfer data at backward angles 
can only be reproduced 
with a spectroscopic factor S$_{t-t}$ which is much smaller than 
the theoretical values and the $\it{t}$-$\it{t}$ configuration 
is necessary for the description of the $^6$He ground state wave function.

%
%
\begin{acknowledgments}
The support provided by the SPEG staff of GANIL during the experiment 
is gratefully acknowledged. 
We would like to thank Y. Blumenfeld and N. K. Timofeyuk for fruitful 
discussions during the course of this work. 
This work was financially supported by the 
IN2P3-Poland cooperation agreement 02-106. 
Additional support from the Human Potential Part of the FP5 European Community 
Programme
 (Contract No HPMT-CT-2000-00180) is also acknowledged.
\end{acknowledgments}


\end{document}